\documentclass{ecai}
\usepackage{times}
\usepackage{graphicx}
\usepackage{latexsym}

\usepackage{todonotes}
\usepackage{color}

\usepackage{comment}

\usepackage[square,sort,comma,numbers]{natbib}

\begin{document}

\title{Principles to Practices for Responsible AI: \newline
Closing the Gap}

\author{Daniel Schiff\institute{Georgia Institute of Technology, United States, email: schiff@gatech.edu}
\and Bogdana Rakova\institute{Accenture, Responsible AI, United States, email: bogdana.rakova@accenture.com}
\and Aladdin Ayesh\institute{De Montfort University, United Kingdom, email: aayesh@dmu.ac.uk}
\and Anat Fanti\institute{Bar-Ilan University, Israel, email: {anat\_fanti}@yahoo.com}
\and Michael Lennon\institute{CAIPP.org, email: lennon@caipp.org}}

\bibliographystyle{ecai}
\maketitle

\begin{abstract}
  Companies have considered adoption of various high-level artificial intelligence (AI) principles for responsible AI, but there is less clarity on how to implement these principles as organizational practices. This paper reviews the principles-to-practices gap. We outline five explanations for this gap ranging from a disciplinary divide to an overabundance of tools. In turn, we argue that an impact assessment framework which is broad, operationalizable, flexible, iterative, guided, and participatory is a promising approach to close the principles-to-practices gap. Finally, to help practitioners with applying these recommendations, we review a case study of AI's use in forest ecosystem restoration, demonstrating how an impact assessment framework can translate into effective and responsible AI practices.
\end{abstract}

\section{Introduction}
Artificial intelligence (AI) is already in use across many areas of social and economic life, and new opportunities for AI to contribute to social good (AI4SG) have also been proposed and developed \cite{cowls_designing_2019}. For example, efforts like Microsoft's AI for Earth program highlight the potential of AI to address the United Nation's Sustainable Development Goals (SDGs). However, many challenges face the practical implementation of AI for social good efforts. Similarly, in the field of fairness, accountability, and transparency of AI, decades of research has only recently begun to be more thoroughly incorporated into practical settings, and many questions remain. In this paper we review challenges in translating principles into practices and propose recommendations towards closing this gap.

After introducing prior work on responsible AI principles and concerns about the practical application of these principles in Section 1, Section 2 proposes five explanations for the principles-to-practices gap. We discuss the complexity of AI's impacts, confusion about the distribution of accountability, a social technical disciplinary divide, identifying and using tools, and organizational processes and norms as key issues in this gap.

In light of these concerns, Section 3 proposes the criteria of a framework that could help organizations turn responsible AI principles into practices. We propose that impact assessment is a promising approach towards meeting these criteria, as it has the potential to be sufficiently broad, operationalizable, flexible, iterative, guided, and participatory. As an exemplar, we focus on the new Institute of Electrical and Electronics Engineering’s (IEEE) 7010-2020 Recommended Practice for Assessing the Impact of Autonomous and Intelligent Systems on Human Well-being (henceforth IEEE 7010). IEEE 7010 is a standard that assesses the well-being implications of AI and employs a well-being impact assessment to do so. Finally, to help practitioners apply these recommendations. Section 4 applies a well-being impact assessment framework to a case study. The case study reviews challenges with AI's use in ecosystem restoration and afforestation -- an important aspect related to several SDGs -- and demonstrates how an impact assessment framework may help to close the principles-to-practices gap in this case. 

\subsection{Principles}
As of 2019, more than 20 firms\footnote{For example, Google, Microsoft, IBM, Sage, Workday, Unity Technologies, and Salesforce} have produced frameworks, principles, guidelines, and policies related to the responsible development and use of artificial intelligence (AI).\footnote{We opt for the phrase `responsible AI' but this topic can also be termed as the `ethical AI,' 'trustworthy AI' or similar.} These documents are meant to address many of the social and ethical issues that surround AI, ranging from labor displacement \cite{autor_why_2015} and algorithmic bias \cite{bolukbasi_man_2016} to privacy, an increasingly important issue in the context of the COVID-19 virus \cite{nanni_give_2020}. These governance documents typically address a set of social and ethical concerns, propose principles in response, and in some cases offer concrete reforms or internal governance strategies. 

Research on the various AI documents produced by firms along with government actors and non-governmental associations has identified clear consensus in organizations' ethical priorities \cite{floridi_unified_2019, mittelstadt_ai_2019, Fjeld_Achten_Hilligoss_Nagy_Srikumar_2020}. The social and ethical concerns highlighted most often surround general concern for public, customer, and employee welfare; algorithmic bias and fairness; transparency and explainability; trust in AI; and reliability and safety of AI products \cite{schiff_whats_2020}. While this scholarship often focuses on identifying consensus across organizations \cite{Jobin_Ienca_Vayena_2019}, it has also examined how companies define their responsibilities \cite{greene_better_2019} and whether there are issues neglected across documents.

Importantly, a key focus of the documents is on presenting a set of high-level principles for responsible AI. For example, Google's AI principles include ``Be socially beneficial,'' ``Avoid creating or reinforcing AI bias,'' ``Be built and tested for safety,'' and ``Be accountable to people,'' among other principles \cite{google_ai_2018}. OpenAI discusses its focus on ``Broadly Distributed Benefits,'' ``Long-Term Safety,'' ``Technical Leadership,'' and its ``Cooperative Orientation'' \cite{openai_openai_2018}. However, these and other high-level responsible AI principles can often be vague, host a multitude of possible interpretations, and may be difficult to translate into everyday practices.



\subsection{Practices}
Many scholars have raised concerns that companies less often provide detailed prescriptions of policies or practices meant to ensure that these principles are adhered to \cite{greene_better_2019, Jobin_Ienca_Vayena_2019, schiff_whats_2020}. In some cases, companies have established relatively clear strategies. Proposed practices include training, hiring, algorithm development frameworks and tools, and governance strategies. For example, Vodafone's AI Framework provides some detail on specific actions it will take, such as adhering to its Code of Conduct and privacy commitments \cite{vodafone_group_plc_vodafones_2019}. SAP proposes as part of its Guiding Principles an AI Ethics Steering Committee and an AI Ethics Advisory Panel \cite{sap_saps_2018}. IBM's Everyday Ethics for AI provides a set of recommended actions and questions for its employees to address key concerns \cite{ibm_everyday_2018}. 

On the other hand, some principles are not accompanied by clear expressions of changes to practice. For example, documents from Tieto, Futurice, and Salesforce focus on abstract principles and commitments. Futurice proposes to ``avoid creating or reinforcing bias,'' while Salesforce claims ``we test models with diverse data sets, seek to understand their impact, and build inclusive teams'' and Tieto states it is ``committed to harness AI for good, for the planet and humankind'' \cite{futurice_futurice_2018, tieto_corporation_tietos_2018, salesforce_salesforce_2019}. These and other generic principles like ``Be socially beneficial`` beg the question of how exactly the companies are carrying out their commitments. 

In the best case, companies may still be in the process of working out the details or may have communicated their intended strategies in other venues, for example, by publishing tools for responsible practice. Nevertheless, remaining at the ``mission statement'' level and the lack of practical detail are worrisome. 
We believe that the question of translating high-level principles into effective and responsible practices is a critical priority in the near-term future for AI. Closing the principles-to-practices gap is worthy of attention by companies developing AI, by those who might procure and deploy AI systems, and by other stakeholders and the public more broadly.

\subsection{Principles without Practices?}

Despite firms' efforts towards establishing some design and development principles for AI systems, several breaches of law and the public trust have been reported in the last few years. Companies have come under significant scrutiny, in some cases facing significant negative media attention, along with customer criticism and employee petitions, walkouts, and resignations \cite{shane_business_2018}.

The question then is why principles have seemingly not been translated into effective practices? In fact, the process of translation is neither obvious nor automatic as clear. According to Mittelstadt (2019) 
``norms and requirements can rarely be logically deduced… without accounting for specific elements of the technology, application, context of use, or relevant local norms''. Barring practical guidance and absent ``empirically proven methods… in real-world development contexts,'' \cite{mittelstadt_ai_2019} claims of responsible AI may amount to no more just that --- claims. 

As a result, some criticisms more deeply impugn the motives of firms. Greene et al. \cite{greene_better_2019} argue that companies attempt to shift responsibility onto designers and experts in order to minimize scrutiny of business decisions. Similarly, Hagendorff \cite*{hagendorff_ethics_2020} argues that companies are often driven by an economic logic, and that ``engineers and developers are neither systematically educated about ethical issues, nor are they empowered, for example by organizational structures, to raise ethical concerns.'' On this account, companies may be strategically promoting their principles to ameliorate customer trust and reputational concerns. In this way, they can appear actively engaged regarding AI's ethical risks in the public eye, but while framing issues so as to minimize genuine accountability.

While some of these deeper criticisms may be true in part or for some organizations, we think a more multifaceted and charitable interpretation \cite{abebe_roles_2020, bietti_ethics_2020} is both appropriate and likely to be beneficial toward seeking positive change. Organizations are best understood not as monolithic single actors, but as multiple coordinating and competing coalitions of individuals \cite{march_business_1962}. Individuals within a single organization may have multiple or competing preferences and roles. Organizational motives should therefore be considered a complex composition of genuine ethical concern, economic logic, signaling and framing strategies, and promotion of both internal and external changes \cite{schiff_whats_2020}. 

Researchers who have noticed the principles-to-practices gap have begun proposing strategies \cite{Cussins_Newman_2020}, often aimed at companies. These proposals include changes to software mechanisms (such as audit trails), hardware mechanisms (such as secure hardware enclaves), and institutional mechanisms (such as red team exercises) \cite{brundage_toward_2020}. This work highlights that it is not only technical practices that must adapt, but also organizational practices.

Among the most comprehensive work assessing the principles-to-practices gap is the review by Morley et al. (2019), which systematically explores existing responsible AI tools and methodologies mapped against six components of the AI development lifecycle: 1) business and use-case development, 2) design phase, 3) training and test data procurement, 4) building, 5) testing, 6) deployment, and 7) monitoring \cite{morley_what_2019}. They identify 106 such tools and methodologies. Some such methods are relatively narrower in scope, such as those surrounding explainable AI \cite{guidotti_survey_2018}, bias \cite{bolukbasi_man_2016}, or procurement (e.g., the AI-RFX Procurement Framework). 

Other methodologies adopt a broader scope of focus, including impact assessments like the ISO 26000 Framework for Social Responsibility \cite{zhao_improving_2018} and IEEE 7010 \cite{musikanski_ieee_2018}. Relevant methods and approaches for responsible AI also come from outside of the AI domain and include privacy-by-design \cite{oetzel_systematic_2014}, value-sensitive design \cite{friedman_survey_2017}, the Responsible Research and Innovation (RRI) approach \cite{schomberg_vision_2013}, and numerous others. In fact, the plethora of possible tools is itself a challenge which we discuss more in Section 2.

\section{Explaining the Principles-to-Practices Gap}

In short, despite the urgent attention to responsible AI in recent years, there are already many existing frameworks and a growing set of new methods aimed at addressing core ethical issues. Why then does the issue of translating principles to practices seem intractable? We offer a few candidate explanations that are neither exhaustive nor mutually exclusive.

\subsection{The Complexity of AI's Impacts}
AI's impacts on human well-being -- positive or negative -- are more complex than is sometimes assumed. Site-based research has identified that engineers are often focused on single products and the physical harm they may cause rather than broader kinds of harms, such as social, emotional, or economic harms \cite{vakkuri_ethically_2019}. Even as conversations surrounding responsible AI increase, most work centers around a relatively small subset of issues, most often bias \cite{bellamy_ai_2018} and transparency \cite{adadi_peeking_2018} in particular AI models. This approach involves exposing and then attempting to mitigate bias in algorithms as well as trying to improve interpretability or explainability given the black-boxed nature of certain AI models which can make decision-making processes opaque. Other commonly-emphasized issues include privacy, reliability, and safety. 

However, these prominent issues most familiar to engineers still constitute only a subset of social and ethical risks and impacts related to AI. Indeed, AI can be understood to impact a wide variety of aspects of human well-being, such as human rights, inequality, human-human relationships, social and political cohesion, psychological health, and more. AI can also impact natural ecosystems and animal life.\footnote{Impacts on the environment and non-human animals may be intrinsically important, as well as instrumentally important to human well-being.} Moreover, many of these harms do not arise in a straightforward way from a \textit{single} AI product, but from many AI systems influencing human social and economic life together and over time.

AI is not the only technology with complex implications on human well-being. Yet its rapid rise is leading to calls for urgency, and some aspects of AI surface a unique combination of ethical concerns \cite{cowls_designing_2019}. For example, compared to other general-purpose technologies like electricity or the internet, AI is notable for its autonomy, its capacity to `learn,' and its power in making accurate predictions, all while embedded in software and ambient systems and therefore invisible to many affected by it. As a result, AI systems are becoming increasingly ubiquitous, and can act in the aggregate to influence human and social well-being in subtle but pervasive ways. 

For example, algorithms on social media designed to steer consumers to entertaining video clips have also led to so-called filter bubbles that may foster political polarization, misinformation and propaganda, targeting of minority groups, and election interference. AI as instantiated in autonomous vehicles has potentially massive implications for physical infrastructure, energy and environment, traffic fatalities, work productivity, urban design, and unemployment \cite{bagloee_autonomous_2016}. In short, addressing AI principles in full seriousness requires an expansive scope of attention to the full set of issues influencing human well-being. This requires looking well beyond a narrow set of topics such as bias, transparency, privacy, or safety and treating them as independent issues. Instead, the full range of topics and their complex interdependencies needs to be understood. However, such a task can be enormously difficult.

\subsection{The Many Hands Problem}
It is clear that responsibly designing and applying AI is therefore both a technical challenge and a social one (implicating social, economic, and policy questions). For example, creating a facial recognition system for policing that minimizes racial bias (by some technical measure) is inseparable from questions on the legitimacy of the use of that system in a particular social and policy setting. However the question of distributing accountability for addressing these issues remains open and contested. Engineers and computer scientists may see their responsibility as focused on the quality and safety of a particular product rather than on larger scale social issues, and may be unaware of the wider set of implications \cite{faulkner_nuts_2007}. Business managers and companies may see their responsibility as fiduciary, in producing high-quality products and revenue. This potentially creates holes in responsibility for addressing key well-being impacts of AI.

In addition to uncertainty regarding one's scope of professional accountability, engineers and computer scientists who focus on design of systems may have limited influence within their organizations. They may expect business managers, liability officers, or corporate social responsibility staff to assess broader social and ethical issues. Social scientists and ethicists tapped specifically for these issues may find themselves similarly handicapped, perhaps in an external advisory role without real say. The result is the `many hands' problem, where responsibility for responsible AI is distributed and muddled \cite{floridi_distributed_2013}. The many stakeholders involved in shaping AI need to be both functionally able and willing to resolve the accountability question with a concrete division of labor. If companies fail to resolve these challenges, they may continue to face public scrutiny as well as financial and legal risks and reputational harms. Moreover, they may harm their employees, consumers, or the public. Figuring out how to distribute responsibility for AI's impacts on well-being is therefore as critical as it is difficult. It may involve challenging long-held assumptions and shifting norms.

\subsection{The Disciplinary Divide}
Another related challenge is the plurality of professional disciplines with roles to play in responsible AI. Discourse on responsible AI has been advanced not only by engineers and computer scientists, but also by sociologists, ethicists, historians and philosophers of technology, policy scholars, political decision-makers, journalists, members of the public, and more. Yet the composition of these diverse stakeholders directs attention to the likelihood that they may bring very different perspectives to the table. They may differ in their technical and ethical education, their framing of problems and solutions, their attitudes and values towards responsible AI, and their norms of communication.

Consider attempts to apply the principle of fairness in attempting to minimize bias. Arguably, a thoughtful AI engineer today might identify a normative principle like 'fairness,' specified in a corporate responsible AI policy, pick a plausible fairness metric to instantiate it (noting there are ineliminable trade-offs between different metrics \cite{chouldechova_fair_2017}), apply it, and communicate these decisions transparently \cite{leben_normative_2020}. However, even these laudable efforts cannot begin to satisfy the extensive societal questions related to fairness, discrimination, and inequality that trouble many social scientists and ethicists. 

More specifically, approaching social issues like bias and fairness too narrowly leads to what Selbst et al. (2018) call category or abstraction errors. For example, computer scientists and engineers developing AI systems can fail to consider how an AI system will be implemented in different social contexts, influence human behavior in those contexts, or lead to long-term ripple effects, all of which can threaten the assumptions on which the AI system is built. This is especially difficult as predicting a technology's usage and impact is known by historians of science and technology to be difficult \cite{winner_artifacts_1980}. More fundamentally, AI developers may err in even considering social concepts like fairness to be computationally definable and technically soluble \cite{selbst_fairness_2018}. 

Consider an algorithm designed to minimize racial bias that is used to inform a judge's decision about criminal sentencing. An algorithm designed and trained on test data from one jurisdiction may translate poorly to another region. It may influence the judge's decisions in unexpected ways, as a judge may overtrust or undertrust the algorithm, or even hold values contrary to those reflected in the algorithm. For example, the algorithm may favor predictive accuracy, while the judge favors leniency and second chances. The consequences for criminal justice outcomes when such a system is used in complex contexts is unclear, and may feed back in unexpected or problematic ways if an AI is trained on data the system has itself helps to generate. To reiterate, there are many questions about responsible AI that cannot be straightforwardly addressed with a narrow technical lens.

On the other hand, social scientists may bring a lens that is broader but faces an inverse problem to the problem faced by engineers. Frameworks for considering social and ethical consequences of AI more in line with the thinking of social scientists can be unhelpfully complex and vague, and therefore fail to translate into action. For example, ethicists recognize that concepts like justice are complex, while political scientists know that values surrounding justice are politically contested. Yet AI engineers must define some measure of justice to implement it. 

In addressing issues like inequality, social scientists may propose large structural changes to economic and social systems, some of which are difficult to achieve (e.g., reforming the motives of corporations) and others possibly far-fetched (e.g., changing the structure of capitalism). These structural changes may be significantly outside of the scope of control of AI engineers. Also unhelpful are conceptions of AI based on sweeping, overly futuristic, or unrealistic generalizations. These abstractions can fail to provide the specificity needed to think clearly about addressing harms to human well-being. Again, while the intentions may be laudable, translating them to practice can be unfeasible or at best unclear. In the best case, it is difficult to resolve the awkwardness of attempting to apply technical fixes to fundamentally socio-technical problems. Something is lost in translation.

\subsection{The Abundance of Tools}
As we have seen, there are already many tools and methodologies for addressing responsible development and use of AI. While creating more and better such tools and methodologies is a worthy pursuit, in one sense there are already too many. Even those tools that do exist have arguably not been tested sufficiently to demonstrate which are most effective and in which contexts \cite{mittelstadt_ai_2019}. An over-abundance problem makes it difficult for individuals to sort through and assess the utility of a given tool, or to weigh it against the many other available tools. People's time, attention, and cognitive capacity is limited, leading to search and transaction cost problems. As a result, individuals and organizations may fail to take advantage of the useful tools and methodologies that are already out there. 

In addition, many tools and methodologies are not supported by practical guidance \cite{morley_what_2019}. A published journal paper or open source code may explain basic functionality but not contain sufficient instructions to apply, customize, or troubleshoot tools and methodologies, especially in a variety of organizational contexts and use cases. This means that only tools that are well-documented, perhaps those created by well-resourced companies or universities and backed up by online communities, may be feasible to use. Individuals without high levels of expertise and specific training may have little luck even with these prominent tools. 

Further, because of the disciplinary divide, methodologies developed in part or in whole by disciplines outside of engineering and computer science (such as responsible research and design ethics) may have a harder time gaining traction. If these extra-disciplinary ideas are not documented and translated for use in AI development settings, there may be little uptake. More work is needed to test tools empirically, to streamline access and guidance, and to help with sorting between tools and methods. Organizations may need an overarching framework to help integrate these lower-level tools and methodologies.

\subsection{The Division of Labor}
The last explanation for the principles-to-practices that we discuss is how organizations structure their job responsibilities and workflow related to AI. Again related to the disciplinary divide, a major concern is that the computer scientists and engineers more directly responsible for an AI system's development may be functionally separated from other workers likely to be tasked with thinking about the system's broader implications -- such as higher-level business managers, the C-suite, and corporate social responsibility and compliance staff. For simplicity, we refer to these crassly as `technical' and `non-technical' teams.

For instance, several companies have proposed external AI ethics advisory or governance boards. External boards (and likely internal ones) may constitute functionally distinct units of the organization that interact only occasionally with primary AI system designers. The same functional separation may apply even when non-technical teams are internal to an organization.

Non-technical employees may have limited ability to understand or modify an AI system's design if interaction with technical teams happens at an arm's distance. Staff without disciplinary expertise in engineering and computer science and even those with technical expertise but not involved in the system's creation may not be able to imagine improvements to the system's development or deployment. They may make underinformed or overly simplistic decisions, for example, prohibiting the use of an AI system that could be modified; or recommending the use of an AI system when they do not fully understand its risks. This functional separation therefore limits their ability to support responsible AI development that adequately considers the full range of impacts on human well-being.

On the other hand, engineers and computer scientists in technical teams may also not be privy to the deliberations of their non-technical counterparts if there is functional organizational separation. If technical employees are exempt from this dialogue, they will not be able to participate in how their colleagues weigh considerations of corporate responsibility, profit, policy, and social and ethical impacts. They may not learn how to incorporate these concepts and trade-offs into their design processes. Technical teams may also fail to imagine ways in which the system they are creating could be improved, or how other systems, tools, or methodologies could be applied to better safeguard and improve human well-being. In sum, functional separation of technical and non-technical experts in organizations limits the potential to communicate effectively, understand issues robustly, and respond to considerations of AI's impacts on well-being.

\subsection{Summarizing the Concerns}
In this section, we have reviewed five sets of concerns that we believe help to explain why AI principles do not easily translate into concrete and effective practices: 1) that AI's social and ethical implications for human well-being are broader, more complex, and more unpredictable than is often understood; 2) that accountability for ethical consequences is divided and muddled; 3) that the orientations of experts in different disciplines lead to emphases that are too narrow, too broad, and generally difficult for translation and interdisciplinary communication; 4) that existing methodologies and tools for responsible AI are hard to access, evaluate, and apply effectively; and 5) that organizational practices and norms which divide technical from non-technical teams minimizes the chance of developing well-considered AI systems that can safeguard and improve human well-being.





\section{Closing the Gap}

\subsection{Criteria of an Effective Framework for Responsible AI}
Given the proposed explanations above, how can we begin to close the principles-to-practices gap? We think an overarching framework for responsible AI development can help to streamline practice and leverage existing tools and methodologies. What would be the desiderata of such a framework for responsible AI?\footnote{Akin to what Dignum calls `Ethics in Design' \cite{dignum_ethics_2018}} As a starting point and based on the identified gaps, we suggest the following:

\begin{itemize}
    \item \textbf{Broad}: it should consider AI's impacts expansively, across many different ethical issues and aspects of social and economic life. Narrower tools and methodologies such as bias mitigation, privacy-by-design, and product design documentation \cite{gebru_datasheets_2020, mitchell_model_2019} can then be subsumed under this more comprehensive framework. For example, after identifying the scope of an AI system's impacts to human-wellbeing, designers could determine which lower-level sub-tools and methodologies are relevant.
    \item \textbf{Operationalizable}: it should enable users to cast conceptual principles and goals into specific strategies that can be implemented in real-world systems. This includes identifying relevant actions and decisions, assigned to the appropriate stage of an AI's lifecycle, e.g., use case conception, system development, deployment, monitoring. This also means identifying accountable parties for these decisions at multiple levels of governance --- engineers, designers, lawyers, executives. This helps to ensure that accountability for actions is assigned to those with the capacity to implement them. 
    \item \textbf{Flexible}: it should be able to adapt to a wide variety of AI systems, use cases, implementation contexts, and organizational settings. A flexible framework has greater applicability to more kinds of AI systems and use cases, allowing for shared language and learning, while enabling sufficiently customization.
    \item \textbf{Iterative}: it should be applied throughout the lifecycle of an AI system and repeatedly as the AI system, implementation context, or other external factors change, not only at one point. Responsible AI is not one-and-done.
    \item \textbf{Guided}: it should be easy to access and understand, with sufficient documentation for users of moderate skill to apply, customize, and troubleshoot across different contexts. It should also be tested in different contexts with evidence of effectiveness made public.
    \item \textbf{Participatory}: it should incorporate the perspectives and input from stakeholders from a range of disciplines as well as  those that may be impacted by the AI system, especially the public. Translating principles ``into business models, workflows, and product design'' will be an ongoing effort that requires engineers, computer scientists, social scientists, lawyers, members of the public, and others to work together \cite{latonero_governing_2018}.

\end{itemize}

A framework that meets these criteria balances the need for technical specificity with an equally important need for conceiving of AI's  impacts on human well-being in their full breadth and complexity, understanding we cannot fully predict all of AI's possible ramifications. That is, while some prominent strategies for responsible AI assume there are only a small set of issues to address, such as bias, transparency, and privacy, we have argued that AI's impacts are more complex. 

We think that impact assessments are a promising strategy towards achieving these criteria \cite{calvo_advancing_2020}. Impact assessments have been used historically in human rights \cite{latonero_governing_2018}, in regulatory contexts \cite{Radaelli_2009}, and more recently to study the impact of AI or algorithms \cite{reisman_algorithmic_2018, calvo_advancing_2020}. We focus specifically on the recently published IEEE 7010 standard as an exemplar \cite{musikanski_ieee_2018, Schiff_Ayesh_Musikansi_2020} created specifically to assess AI's impacts on human well-being.\footnote{IEEE is in the process of developing standards surrounding ethical AI on a variety of topics --- bias, privacy, nudging, trustworthiness of news, and overall human well-being \cite{Chatila_Havens_2019}. While the authors of this paper were involved in helping to develop IEEE 7010, this paper reflects the individual views of the authors and not an official position of the IEEE.} We argue below that impact assessments like the well-being impact assessment from the IEEE 7010 standard could be adopted by companies pursuing responsible AI development as well as incorporated into the institutions which train future practitioners.

\subsection{Impact Assessments for Responsible AI}
Building on the IEEE 7010 standard, a well-being impact assessment is an iterative process that entails (1) internal analysis, (2) user and stakeholder engagement, and (3) data collection, among other activities. Internal analysis involves broadly assessing the possible harms, risks, and intended and unintended users and uses of an AI system. Here, developers and managers of an AI system carefully consider a wide range of an AI system's potential impacts on human well-being, not limited to prominent issues like privacy, bias, or transparency. 

Critically, assessing impacts requires not just speculating about impacts, but measuring them. Therefore, the user and stakeholder engagement stages of the assessment include learning from users of AI systems as well as others more indirectly impacted to determine how the system impacts their well-being. When developers have access to users, this may include asking them about possible or actual psychological impacts, economic impacts, changes to relationships, work-life balance, or health. . This is again in contrast to strategies which focus solely on technical fixes to issues like bias or privacy during the design stage alone and fail to account for the broader universe of well-being implications.

Finally, data collection based on the continuous assessment of the identified possible impacts is key. Here we refer to the collection and tracing of data related to the impact assessment, which may exceed collection of data related to the development of the AI system itself. Data can be collected through user surveys, focus, groups, publicly-available data sources, or directly as system outputs. In sum, we propose that adherence to -- and rigorous documentation of -- impact assessments will contribute to the continuous improvement of AI systems by ensuring that organizations are better able to understand and address AI's many impacts on human well-being.

Not all tools entitled `impact assessment' meet our definition. Many existing tools consider only a small scope of possible impacts. Some fail to {measure} impacts at all, instead focusing on anticipating impacts assumed to be important and applying best practices to avoid associated harms.  
Inversely, some tools that are not labelled `impact assessments' might be classified as such under our definition, such as the European Commission's Ethics Guidelines for Trustworthy AI \cite{European_Commission_2019}. Notably, some frameworks have been proposed by the public sector (i.e., governments) and others by non-governmental organizations and companies.

Why is impact assessment as we have defined it a promising approach? First, it can be highly \textbf{broad}, measuring many aspects of human social and economic well-being and even environmental impacts. IEEE 7010 is an exemplar in its breadth. It identifies twelve domains as part of its well-being impact assessment: affect, community, culture, education, economy, environment, health, human settlements, government, psychological/mental well-being, and work \cite{musikanski_ieee_2018}.\footnote{All discussion of IEEE 7010 is adapted and reprinted with permission from IEEE. Copyrights IEEE 2020. All rights reserved.} For an AI system like a chatbot or autonomous vehicle, the impact assessment may lead to identification of numerous areas of concern like social relationships, the environment, psychological health, and economy. Thus while other responsible AI approaches take into account a far narrower range of concerns, emphasizing largely bias and transparency of algorithms, IEEE 7010's well-being impact assessment is far more broad-ranging. 

Impact assessments can also be highly \textbf{operationalizable} into specific strategies. In IEEE 7010, overarching domains like human well-being or environmental impacts are not just stated in abstract terms, but are measured through specific indicators based on rigorous research. The strategy involves internal analysis, followed by user and stakeholder engagement, both used to determine domains where an AI system can impact human well-being. Next, AI system creators can identify measurable indicators related to each domain, followed by measuring the impacts of their AI system on the selected indicators. For example, through using the well-being impact assessment, an AI developer might identify the environment as an important concern, and increases in air pollution as a specific possible impact. Using validated indicators to measure air pollution, the developer could then assess whether air pollution has increased or decreased. 

Next, given the extensive range of possible impacts that can be measured, there is also ample room to customize an impact assessment and make it sufficiently \textbf{flexible} for particular use cases and organizational contexts. ALGO-CARE is one such example of an impact assessment applied specifically to algorithms in the criminal justice system \cite{oswald_algorithmic_2018}. ALGO-CARE considers context-specific issues like whether human officers retain decision-making discretion and if proposed AI tools improve the criminal justice system in a demonstrable way. Similarly, users of IEEE 7010 would find that their impact assessment approach could be customized to focus on issues ranging from housing to human rights. Impact assessments also typically leave room to determine which actions are taken in response to identified impacts, meaning these responses can be applied in an \textbf{iterative} fashion, not only during the design phase. For example, concerns about impacts of an AI system on pollution could lead to changes not only during an AI system's design, but also in terms of its implementation in the real world. Moreover, it is impossible to finish an impact assessment only during the design stage, as it requires measuring its impacts in real-world settings.

However, this breadth and flexibility suggest to us that \textbf{guidance} is the most challenging issue currently. Simply, there is no one-to-one mapping from identified impacts or problems with AI systems to individual technical or implementation `fixes' and creating such a comprehensive mapping is likely not plausible. The breadth and complexity of an AI well-being impact assessment demonstrate the difficulties for any actor who attempts to single-handedly close the gap from principles to practices. Thus, we propose that developing guidance for particular sectors, types of AI systems, and use cases is a necessary and ongoing effort which could leverage a \textbf{participatory} process-driven impact assessment approach that engages different groups of stakeholders.\footnote{Importantly, the impact assessment tool selected need not be IEEE 7010's well-being impact assessment; impact assessment frameworks with developed by the Canadian government \cite{canada_2019} and AI Now \cite{reisman_algorithmic_2018} are examples of other promising tools already available.} In particular, developers, policymakers, philosophers, intended and unintended users of the technology being developed, and others could equally contribute in the AI impact assessment process, such as through interviews, focus groups, and participatory design methods \cite{Schuler_Namioka_1993}. 


We are hopeful that more scholars and organizations focused on responsible uses of AI will adopt an assessment approach that measures a wide range of impacts on human well-being and meets the criteria identified above. A key aspect of creating the supportive structure for effective impact assessments will be adopting new educational practices in institutions of higher education as well as organizational changes in firms. We turn to these issues briefly.

\subsection{Supportive Practices in Institutions of Higher Education}
Educational institutions also have an important role to play. Educational systems have undertaken meaningful efforts aimed at increasing ethical sensitivity and decision-making, but have not yet made the changes needed to support responsible AI practice. Of around 200 AI/ML/data science courses reviewed by Saltz et al. (2019), little more than 1 in 10 mentioned ethics in their syllabus or course description. Those that did focused overwhelmingly on bias, fairness, and privacy \cite{saltz_integrating_2019}. While courses focused specifically on AI ethics cover a wider set of issues including consequences of algorithms, technically tractable issues like bias and privacy are still prominent \cite{garrett_more_2020}. We suggest that AI ethics education focus not solely on a few prominent or technically tractable issues nor on general awareness building alone, but also on impact assessment as an overarching framework to understand AI's impacts on human well-being.

AI ethics and design courses should also recruit and serve students of social sciences and humanities (and other `non-technical' fields). Calls for more STEM education for these individuals often result in them taking a small number of basic computer science or statistics courses. We believe that more fundamental interaction with AI systems is important to build capacity in these students, who should be ``capable of grasping technical details'' in order to translate abstract principles and concepts from these fields into concrete computer and data ethics practices \cite{hagendorff_ethics_2020}. In turn, students in social scientists and humanities can help to expand the scope of thinking of their counterparts in engineering and computer science. For example, AI ethics and design courses can facilitate interdisciplinary teamwork that involves the use of impact assessments. Such an approach would allow students to understand the range of AI's impacts and practice applying relevant tools and methodologies in response. Interdisciplinary teaming could also occur through student extracurricular clubs and contests (not limited to grand prizes) to encourage this kind of cross-disciplinary learning and practice.

\subsection{Supportive Practices in Business Organizations}
Analogous to the educational setting, companies developing or deploying AI should move towards the integration of technical and non-technical teams rather than functional separation of roles, for reasons discussed in the previous section. These integrated teams could include technical developers as well as other individuals tasked with considering impacts of an AI system who may have social science, humanities, business, law, or ethics expertise, or who can represent a typical user's perspective effectively. Such a change requires establishing practices that are integrated with engineering and software lifecycles and part of the ongoing dialogue characteristic of development processes. Already, organizations have proposed including a residential non-technical thinker tasked with responsible AI --- an `ethics engineer' or 'responsible AI champion' \cite{OBrien_Sweetman_Crampton_Veeraraghavan_2020}. 

However, we would urge that these integrated teams not remain at an arm's distance in a way that maintains bifurcated expertise areas and roles. Instead, technical and non-technical team members should aim learn each other's languages and work jointly. For example, an AI development team could include ethnographers, policy scholars, or philosophers, all tasked with applying a broad impact assessment as the AI system is being created and implemented. While these changes to organizational practice may be difficult, requiring individuals to stretch their boundaries, we believe that a deep level of integration is necessary to bridge the disciplinary divide. 

Organizations could also engage in interdisciplinary and interdepartmental cross-training, potentially supported by responsible AI champions or external experts. For example, organizations could facilitate red team exercises \cite{brundage_toward_2020} or hypothetical case studies that draw on the impact assessment approach. Practicing even on hypothetical cases allows social science-oriented practitioners and technically-oriented practitioners to learn from one another about how they can define problems, consider solutions, define terminology, etc. This can help diverse disciplinary practitioners begin to learn and establish common language and identify gaps and opportunities in each other's practice.

In summary, we have argued that impact assessments are a promising strategy to address the gaps between principles and effective practices for responsible AI. However, applying an impact assessment might feel like an abstract exercise to those who have not done it. To demonstrate how closing the principles-to-practices gaps with an impact assessment might occur, we move now to a case study.

\section{Case Study: Impact Assessments to Support Responsible AI for Forest Ecosystem Restoration}


In this section, we set out to explore how the recommendations introduced above could be implemented within a particular setting. We hope this case study will help practitioners in adapting our research findings to the unique sociotechnical context within which their own work is situated. In the example case study below, we look at AI systems that are being used to address forest ecosystem restoration. 

\subsection{Case Study Background} 
As is characteristic of the SDGs, achieving goals in one area -- like the environment -- also has effects on multiple other goals, such as addressing health and poverty targets. Forest restoration is one such aspect of the SDGs. While it has clear importance to SDG 13 (Climate Action) and SDG 12 (Responsible Consumption and Production), forest ecosystem restoration is addressed most directly by SDG 15 (Life on Land). SDG 15 states a global ambition to ``Protect, restore and promote sustainable use of terrestrial ecosystems, sustainably manage forests, combat desertification, and halt and reverse land degradation and halt biodiversity loss'' \cite{United_Nations_2020}. 

Forest ecosystem restoration is therefore essential for many reasons. Forests have the most species diversity on the planet, with some 80\% of land-based species. Forests also reduce the risk of natural disasters such as floods, droughts, and landslides and help protect watersheds \cite{United_Nations_2020}. Further, forests are critical for mitigating land-based carbon emissions by increasing carbon sequestration, critical for climate change prevention goals \cite{Nilsson_Schopfhauser_1995}. Project Drawdown, for example, has calculated that the restoration and protection of tropical forests could lead to 61.23 gigatons of carbon reduction by 2050 \cite{Project_Drawdown_2020a}.

Achieving these goals requires the restoration of forest ecosystems through the cultivation of trees, known as afforestation \cite{Project_Drawdown_2020b}. Applied afforestation projects typically involve three stages - planning, execution, and monitoring of ecosystem restoration. Several AI technologies have been used in afforestation efforts and their use is increasing. During planning, AI systems have been used to predict forest carbon sequestration potential through the use of satellite and drone image data \cite{RASOL2007, Obaidy2015}. AI can also facilitate execution of afforestation through computer vision algorithms used in identifying appropriate planting sites, monitoring plant health, and analyzing trends \cite{BioCarbon}. Lastly, in the monitoring stage of restoration projects, AI can be used to identify where deforestation may have been conducted illegally \cite{hethcoat2019machine, lippitt2008mapping}, as well as assess risks due to fire, disease, insects, or other causes \cite{Schmoldt_2001}.

\subsection{Current Challenges}
While AI thus has great potential to contribute to SDG efforts and social good in this case, there are complications with translating the aforementioned goals into responsible practices. We focus on one specific challenge leading to a gap in responsible AI practice -- the issue of multi-stakeholder coordination. 

According to international governance efforts like the UN SDGs, the UN Forum on Forests, Agenda 21, and the Future We Want (the outcome document of the Rio+20 Conference) there is a need for holistic, multi-stakeholder engagement to address forest ecosystem restoration adequately \cite{Sitarz_1993, Carolee_Heather_2014}. This is due to the existence of multiple groups with critical interests in forest ecosystems. Local communities and businesses may engage in harvesting timber, farming, and industrial exploitation to produce resources and support local economies. Natives living off the land have an essential stake, as their livelihood may depends on hunting animals and harvesting plants and other materials. Government officials tasked with maintaining forests or woodlands need to monitor the quantity and kind of trees to harvest, and NGOs focused on conservationism may attend to animal life and biodiversity as well. Finally, policymakers must also worry about carbon sequestration and climate change efforts. 

Though the goals of these groups are not always in conflict, they can come from different perspectives and have competing priorities. Therefore, AI-driven systems used for afforestation that do not take into account these ``multiple ecological, economic, social and cultural roles'' important to various stakeholders \cite{United_Nations_2020} may lead to blind spots and unintended harms. For example, an AI system that uses imaging data to determine carbon sequestration potential could optimize climate change goals in a narrow sense, but fail to account for social-ecological aspects of the land important to indigenous groups, or ignore endangered species important to conservationists. This could engender a lack of coordination and collaboration among stakeholders and lead to costly delays and conflict, as parties are unwilling to accept afforestation efforts or even work actively against them. 

As a result, carbon sequestration targets optimized in the short term could fall short in the long term as afforestation progress fails to translate into a sustainably managed multi-stakeholder effort. Failing to develop and implement AI systems for ecosystem restoration in a participatory fashion is thus an example of how the laudable goal of improving environmental well-being can fail to translate into responsible and effective practices.

\subsection{Applying Impact Assessment}
It is therefore important for developers of AI systems to consider that numerous groups have stakes in forest ecosystem restoration. As discussed by Rolnick et al. in the case of AI \cite{rolnick2019tackling}, ``Each stakeholder has different interests, and each often has access to a different portion of the data that would be useful for impactful [machine learning] applications. Interfacing between these different stakeholders is a practical challenge for meaningful work in this area.'' Landowners, policymakers, public and private sector organizations, local communities, and others need to have a voice in the application of AI to forest ecosystem restoration. 

How would AI impact assessments such as the IEEE 7010 well-being impact assesssment help in this instance? As discussed in Section 3, the assessment process involves a \textbf{broad} internal analysis by the organizations developing AI systems for forest ecosystem restoration. This would involve trying to understand the variety of possible stakeholders and intended or unintended impacts of their products. A company that develops AI to identify target areas for afforestation given carbon sequestration potential might recognize possible impacts on species diversity, the local economy, and the general well-being of native groups.

In order to have a more accurate picture -- as well as to build consensus among stakeholders -- the company would then begin the user and stakeholder engagement process. This would involve talking to local governments procuring the company's AI systems about the need for a holistic implementation of afforestation efforts. Critically, it would involve soliciting the input of the numerous stakeholders mentioned such as conservation groups, landowners, scientists, government officials, local businesses, and native populations. For example, a method like participatory action research or other \textbf{participatory} design methods \cite{Schuler_Namioka_1993} could be used to facilitate this engagement.

This process, which should be ongoing and \textbf{iterative} throughout the management of the forest ecosystem, should surface a number of clear concerns about possible implications of the afforestation efforts. For example, the company may have originally been optimizing a target through their AI system such as SDG indicators 15.1.1, "Forest area as a proportion of total land area," or 15.3.1, "Proportion of land that is degraded over total land area." However, the impact assessment process should lead to the \textbf{flexible} identification of new indicators critical to having a broader understanding of the broader social, economic, and ecological context.  

These new indicators -- reflecting economic, health, and governance concerns as well as environmental ones -- could include, for example, SDG indicators 3.3.5, "Number of people requiring interventions against neglected tropical diseases," 1.5.2, "Direct disaster economic loss in relation to global gross domestic product," or 11.3.2 "Proportion of cities with direct participation structure of civil society in urban planning and management that operate regularly and democratically." These and other indicators, not necessarily picked from the SDG indicators, would therefore \textbf{operationalize} possible dimensions and impacts of the forest ecosystem management effort -- disease, natural disasters, and participatory governance -- as specific \textit{measurable} indicators. The company in collaboration with partners would endeavor to measure these impacts, not merely carbon sequestration or forest area as a proportion of land area.

Finally, the company in collaboration with partners, would have several ways to use this new and deeper understanding of the well-being implications of their AI system. One such approach would be embedding this expert domain knowledge garnered from the participatory process into the architecture of the AI system itself \cite{Kursuncu_Gaur_Sheth_2020}. For example, an AI system that previously optimized carbon sequestration potential as part of its objective function could incorporate new data regarding tropical diseases or natural disasters as additional constraints or targets in the optimization of its model.

However, not all efforts to address the identified well-being impacts need be strictly technical in nature. Changes to organizational practices and governance strategies are likely called for. For example, the company might find that accounting for species diversity directly within the model is not sufficiently nuanced. Instead, the company could bring initial recommendations about carbon sequestration target areas to a multi-stakeholder governance board. The board could then offer feedback on the suitability of the recommendations given species diversity or native land usage. While the impact assessment process and identification of solutions would initially feel unfamiliar and complex, the company would gradually develop best practices and \textbf{guidance} towards a more responsible application of its AI system for forest ecosystem restoration.

While this case study -- the use of AI for forest ecosystem restoration -- is based on real uses of AI and associated real-world challenges, the specific indicators and actions taken by the company are hypothetical. We do not mean to suggest that there are not companies or governments already taking thoughtful approaches to multi-stakeholder governance in this area. However, to the best of the authors' knowledge, current sustainability efforts have not yet incorporated impact assessments of AI-driven technological solutions applied to ecosystem restoration. We hope this case study helps demonstrate how impact assessments are a promising tool to close the principles-to-practices gap towards responsible AI.

\section*{Conclusion}
In this paper, we reviewed and synthesized explanations for the gap between high-level responsible AI principles and the capacity to implement those principles in practice. We identified five explanations for the gap, related to the complexity of AI's impacts on well-being, the distribution of accountability, socio-technical and disciplinary divides, a lack of clarity and guidance around tool usage, and functional separations within organizations that preclude effective interdisciplinary practices. 

Next, we considered the criteria of a framework likely to help close the principles-to-practices gap, and identified that impact assessment is one such approach. An impact assessment approach to responsible AI, unlike some alternative approaches, has the potential to be broad, operationalizable, flexible, iterative, guided, and participatory. After reviewing the benefits of impact assessment and the well-being impact assessment approach of IEEE 7010, we suggested changes that educational institutions and companies can make to supportive effective and responsible AI practices.

Finally, we considered the use of AI in forest ecosystem restoration efforts. In the face of complex impacts and possible conflicts between stakeholders that could inhibit sustainable forest management efforts, impact assessment offers promise for those wishing to close the principles-to-practices gap.


\bibliography{ecai}
\end{document}